# Electric field–induced phase transition and electrocaloric effect in PMN-PT

H. H. Wu[1] and R. E. Cohen[1, 2*]

[1]Department of Earth and Environmental Sciences, Ludwig-Maximilians-Universität, Munich 80333, Germany

[2]Extreme Materials Initiative, Geophysical Laboratory, Carnegie Institution for Science, Washington, DC 20015-1305, United States

## Abstract

Ferroelectric perovskite oxides possess a large electrocaloric (EC) effect, but usually at high temperatures near the ferroelectric/paraelectric phase transition temperature, which limits their potential application as next-generation solid-state cooling devices. We use classical molecular dynamics to study the electric field–induced phase transitions and EC effect in PMN-PT ($PbMg_{1/3}Nb_{2/3}O_3$-$PbTiO_3$). We find that the maximum EC strength of PMN-PT occurs within the morphotropic phase boundary (MPB) region at 300 K. The large adiabatic temperature change is caused by easy rotation of polarization within the MPB region.

## Introduction

The electrocaloric (EC) effect is the adiabatic temperature change (ATC) of a material in response to an applied electric field in dielectric materials. Perovskite ferroelectrics possess a substantial EC effect, which can be used in solid-state cooling devices for a broad range of applications such as on-chip cooling and temperature regulation for sensors or other electronic devices.[1-3] EC materials attracted scientific interest in the 1960s and 1970s.[1,4,5] At that time, the EC effect was not sufficiently high for practical applications because low electric fields induced only a small change in ATC ($\Delta T < 2$ K). In 2006, a giant EC effect of $\Delta T_{max} = 12$ K, nearly an order of magnitude higher than before, was observed in $PbZr_{0.95}Ti_{0.05}O_3$ thin films.[6] Since then, giant EC effects have been reported for various ferroelectric ceramic and polymer films.[7-11] Although the EC effect in thin films is significantly enhanced by applying ultrahigh fields of hundreds kV/cm, the EC strength $\Delta T/\Delta E$ is not improved compared with the bulk counterparts.[12] A large polarization change is necessary to achieve a large EC effect, and hence, switching from ferroelectric phase to paraelectric phase near the paraelectric/ferroelectric transition Curie temperature results in the largest EC effect. [10,13,14] In this study, we examine the EC in PMN-PT as a function of temperature, composition, and applied field. We also model for the first time the phase diagram under applied electric field, previously studied experimentally.[15-17]

---

[*]Corresponding author. E-mail address: rcohen@carnegiescience.edu

By employing classical molecular dynamics (MD) with a first-principles-based shell model potential, Rose and Cohen[14] found a large ATC $\Delta T \sim 25$ K in lithium niobate (LiNbO$_3$) under $\Delta E = 75$ MV·m$^{-1}$ around the paraelectric/ferroelectric transition temperature and suggested that the operating temperature for refrigeration and energy applications should be above the zero-field phase transition temperature T$_c$ to obtain a large electrocaloric response. However, T$_c$ in many ferroelectric materials is considerably higher than room temperature, which substantially limits their potential application in solid-state cooling devices. In the past decade, considerable efforts have been exerted to achieve high EC effects at room temperature,[13,18-26] such as applying a stress field,[20,21,27] doping,[22] introducing defects,[18,23,27] using the tetragonal-cubic phase transition, [10,12] and taking advantage of the morphotropic phase boundary (MPB) region in solid solutions.[13,24,28]

PMN-PT is a solid solution of the relaxor PMN and the ferroelectric PbTiO$_3$. It contains the perovskite structure ABO$_3$ with lead on the A-site, and Mg$^{2+}$, Nb$^{5+}$, and Ti$^{4+}$ occupy the B-sites. An MPB region exists from approximately 0.3–0.4 mole fraction PbTiO$_3$, in which the electromechanical response is highest.[29] The behavior of this MPB region has been understood in terms of the easy rotation of the polarization direction within monoclinic structures. Some previous works have reported the maximum EC effect within the MPB region.[24,28] The mechanism of the EC enhancement near the MPB region is associated with an easy path for polarization rotation. First-principles calculations,[30-34] atomic-level MD simulations[35-38], Monte-Carlo simulations[39] and phenomenological Landau theories[40,41] have been used to study the electric field–induced phase transition in ferroelectrics. Among these methods, MD simulations with shell models fitted to first-principles calculations are sufficient for predicting the behavior of pure compounds and solid solutions.[35,37,38] These potentials account naturally for the presence of many factors such as chemical order or compositional heterogeneity at the atomic level. In the shell model, [42,43] atomic polarizability, crucial for reproducing ferroelectricity in the simulations, is modeled mechanically, where each atom has a charged shell attached to a charged core by a spring. Interatomic interactions come from the charges, as well as a pairwise potential between shells.

An important issue in PMN-PT is the treatment of order at the perovskite B-site, which can be occupied by $Nb^{5+}$, $Mg^{2+}$, or $Ti^{4+}$. Disorder and resulting random fields are essential in relaxor behavior, and the current model that fits the existing experimental data in PMN is the presence of chemical ordered regions (CORs) immersed in a disordered matrix.[44-46] CORs are described by the "random site model" (RSM)[47] in which B-cations display a rock salt type of order, with one sublattice occupied completely by $Nb^{5+}$ and the other containing a random distribution of $Mg^{2+}$ and $Nb^{5+}$ in a 2:1 ratio. Starting from PMN, a sequence of configurations with different Ti content ($x$=0-1.0) was obtained by random substitution of $Nb^{5+}$ and $Mg^{2+}$ by $Ti^{4+}$, preserving the neutrality of the simulated system. We performed simulations for three such sequences (termed 'layouts' below) with different initial PMN configurations to analyze the influence of local order on the ferroelectric properties.[48]

In the current work, we use a shell model fitted to first-principles calculations to study the electric field–induced phase transition and the EC effect at room temperature with respect to concentration and electric field magnitude in a PMN-PT solid-solution single crystal. This model was previously used to predict the elastic constants and the phase diagram of PMN-PT without an external electric field, and the simulation results from this model are in good agreement with the experimental observations. In the present study, we show that this model can also describe the electric field–induced phase transition and the EC effect of PMN-PT over the entire composition range. We also discuss the optimized EC strength at selected compositions. Our findings not only expand the spectrum of room-temperature lead-free EC materials for future refrigeration applications but also may serve as a guide for revealing other EC material alternatives by looking for solid solutions with MPB region.

## Computational methods

Each atom in the shell model is described as two charged and coupled particles: one is a higher mass core, and the other is a lower mass shell. The model also includes electrostatic interactions among cores and shells of different atoms, and short-range interactions between shells. In this work, we use a shell model, where the core and shell are linked through an anharmonic spring, $F(u) = \frac{1}{2}k_2 u^2 + \frac{1}{24}k_4 u^4$, where $u$ is the core-shell displacement. There is an additional penalty term $D_0(u - u_0)^2$ if $u \geq u_0$, where $u_0 = 0.2$ Å and $D_0 = 10000$ ev, to the core-shell coupling to prevent the shell from drifting off the core and ensure atomic potential stability. A short-range interaction between two different atoms describes both the electron cloud repulsion and the van der Waals attraction. We use a Rydberg potential, $V(r) = (A + Br)e^{-r/\rho}$, with a shifted-force correction and a cutoff radius of 10 Å for the short-range interaction between the A–O, B–O, and O–O pairs. The input data to adjust the parameters were from first-principles results within the local density approximation.[49] The model parameters were determined by simultaneous least-squares fitting of the end members PMN and PT.[50] The model parameters used in this work are listed in Table 1.

We use the potential to determine the relaxed structures and finite temperature properties of PMN-PT as functions of composition and external electric field. We use the program DL-POLY classic within the constant (N; σ; T) ensemble at intervals of 10% in concentration.[49] All MD simulations are carried out using with system sizes of 12 × 12 × 12 unit cells (8,640 atoms) under periodic boundary conditions. The thermostat and barostat relaxation times are set to 0.25 ps and 0.35 ps, respectively. All shells are assigned a mass of 2 a.u.[51] The relaxed structures were determined as zero-temperature-limit MD simulations. Our MD runs consisted of at least 100,000 time steps, with data collection after 60,000 time steps, with a time step of 0.4 fs, giving run times of 40 ps. The detailed polarization calculation method can be found in our previous work.[48] The ATC $\Delta$T as the applied electric field change from an initial value of $E_\alpha$ to a final value of $E_\beta$ is calculated from the following indirect method[52]:

$$\Delta T = -\int_{E_a}^{E_b} \frac{TV}{C_{p,E}} \left( \frac{\partial P}{\partial T} \right)_E dE,$$

where $C_{p,E}$ is the specific heat capacity per unit volume and is approximately a constant as $3.093945 \times 10^6$ J/m³K according to the Dulong–Petit law for all the compositions.

Results and discussion

To study the phase transition (or polarization rotation) within the tetragonal phase, the [100] direction response for the [001] direction-polarized tetragonal phase is shown in Figure 1.

To understand the polarization stability regions and the phase transition behavior under a finite electric field for all the compositions of PMN-PT, three types of monoclinic phases, namely, $M_A$ phase $\left(P_1 = P_2 < P_3\right)$, $M_B$ phase $\left(P_2 < P_1 = P_3\right)$, and $M_C$ phase $\left(P_3 > P_1 \neq 0, P_2 = 0\right)$, as well as the tetragonal (T), rhombohedral (R), and orthorhombic (O) phases, are defined in Figure 2. As presented in our previous work,[48] the range of composition from $x = 0.3$ to $0.5$ separates the region of R symmetry from that of T symmetry. For the T structure on the right of the MPB region, namely, $x > 0.5$ in PMN-PT, only one phase transition exists from the T [001] to C phase transition under an external electric field along the [001] direction. Note that this is an isosymmetric transition, as the C phase is tetragonal under an applied electric field. All configurations in the entire composition range become C at high-enough temperature. In the MPB region, various phases of different symmetries are present, which can be distinguished by cell distortions and polarization orientations.

Three layouts[48] with different initial PMN configurations are used to analyze the influence of local order on the properties (Figure 3). When the slope change of the polarization component and lattice constant versus temperature are both larger than 10%, we treated the corresponding temperature approximately as a phase transition temperature. Without an external electric field, the polarizations (Figure 3[a1]) and lattice constants (Figure 3[b1]) of PMN-0.3PT are equal within error at all temperatures. Thus, we find that the phase structure is the R phase when the temperature is lower than 450 K, and transforms to the C phase at higher temperatures. The results under the external electric field $E_3 = 25$ MV/m are shown in Figures 3(b1) and 3(b2). The structure below 400 K is the $M_A$ phase, but it changes to the T phase when the temperature increases. The polarization component $P_i$ and lattice constant under external electric field $E_3 = 50$ MV/m are shown in Figures 3(c1) and 3(c2), respectively. With an increase of the external electric field from 25 MV/m to 50 MV/m, the phase transition temperature from $M_A$ to T is reduced from 400 K to 300 K.

We perform classical MD, so we do not expect correct behavior at low temperatures. Nevertheless, we report the results for completeness. Differing from the composition on the left of MPB region, another composition $x=0.4$ within the MPB region is chosen to better understand the phase transition behavior. Without an external electric field, the three polarization components and the three lattice constants are basically equal for temperatures below 100 K, thus giving an R-type structure (Figure 4). $P_3$ and lattice constant $c$ increase and then decrease with increase in temperature, whereas the polarization component $P_1/P_2$ and lattice constant $a/b$ decrease so that the solid solution transforms to the $M_A$ phase. Finally, the three components of $P_i$ approach zero at $T_C = 550$ K, consistent with the C phase. The sharp change in slope of the polarization versus temperature indicates the transition, although whether it is an isostructural phase transition or a crossover cannot be determined for a finite supercell with a finite sampling of temperatures. Interestingly, the polarization component $P_3$ increases as temperature increases at low temperatures, whereas the total polarization $P_t$ always decreases with increasing temperature. This is due to the polarization rotation mechanism.[31,53]

In the absence of an external electric field, the phase diagram compiled from all of our results correctly reproduces the four phase regions observed in experiments as a function of temperature and compositions, which is consistent with our previous work (Figure 5).[48] With an applied electric field, $E_3$ along the perovskite [001] direction, the phase boundaries move to higher temperatures. More interestingly, the phase transition path under a constant electric field varies with the chang5e of temperature and composition. Take $E_3$=25MV/m as an example, the phase transition path is $M_A \rightarrow$ T$\rightarrow$C for the Ti content $x$<0.45, whereas the phase transition path changes to T$\rightarrow$C when the Ti content is larger than 0.5. Furthermore, the phase transition temperature within the same phase transition path would vary with different composition. The electric field, temperature, and composition dependence of phase transition suggests that the EC effect of relaxor ferroelectric PMN-PT can be effectively tuned by the external electric field and suitable composition engineering.

The reversibility of ferroelectric properties around the phase transition temperature, especially first order phase transition, is a big issue for practical applications. However, relaxor ferroelectrics have diffuse transitions so should have lower hysteretic losses. As presented in Figure 6(a), the polarization component $P_3$ basically increases and then decreases with the temperature increase for all the external electric fields. For a constant external electric field in low temperature range, the polarization component $P_3$ increase with temperature increase. As indicated in Figure 5, the increase of polarization component $P_3$ occurs within the $M_A$ phase, and it reaches a maximum value at the phase transition temperature from $M_A$ to T phase. With further increase of the temperature, the polarization component $P_3$ within the T phase always decreases with the temperature increase. Interestingly, the increase of polarization component $P_3$ occurs at low-temperature regions, and the peak under a constant electric field shifts to a lower temperature with the increase of the external electric field. The increase of polarization component $P_3$ with temperature increase contributes to the negative EC effect[54,55]. The negative EC effect occurs in the temperature range lower than 300 K, whereas the positive EC effect dominates in the higher temperature range (Figure 6[b]). Importantly, the peak of the absolute value of either the positive or the negative EC effect increases with the increase of the external electric field. The simulation results of electric field up/down to 100 MV/m (Figure 6[c]) and temperature up/down to 300 K (Figure 6[d]) show good reversibility, especially at high electric field and/or temperatures.

A primary goal of this study is to find the optimal compositions for electrocaloric applications of PMN-PT at room temperature. We find that $\Delta T$ increases monotonically with applied electric field over all the compositions (Figure 7[a]), as expected.[56] The magnitude of temperature change $\Delta T$ varies for different compositions so that compositions within MPB ($x = 0.3$–$0.5$) show the highest EC effects. Taking $\Delta T$ at $E_3 = 200$ MV/m as an example, the EC effect $\Delta T$ increases with the Ti content increase from $x = 0$ to $x = 0.45$, and then the EC effect $\Delta T$ decreases with further increase of Ti. This result agrees with the experimental observation that the EC effect can be optimized in the MPB region, which was explained by the easy polarization rotation under a finite electric field in this region.[13,19] In fact, it is surprising that the effect is not larger. In LiNbO$_3$, $\Delta T$ varies by 400% for a 15% change in T$_C$ with applied field.[14,57]. In LiNbO$_3$ there is a strong first-order phase transition with a large polarization dropping to zero at Tc in field-free conditions, giving a resultant large $\Delta T$ in the ECE. This is entirely a collinear effect due to crystal symmetry, i.e. field and polarization are aligned. In the MPB region of PMN-PT, polarization rotation is important,[31,53] and the ECE derives from the ease of polarization rotation in the MPB. Furthermore, when the ECE is measured as $\Delta T$, higher temperatures at T$_c$ in LiNbO$_3$ themselves promote higher ECE, for the following reason. Entropy is Int dT C/T, where C is the heat capacity, so at higher temperatures, a smaller entropy change is required to give a given EC effect measured in $\Delta T$. In other words, since large change in polarization always occurs near T$_c$, $\Delta T = -\int_{E_a}^{E_b} \frac{TV}{C_E} \left( \frac{\partial P}{\partial T} \right)_E dE$ , even if $\left( \frac{\partial P}{\partial T} \right)_E$ and the external electric field E range are the same, higher Tc gives higher ΔT. On the other hand, it is the entropy change that governs the cooling efficiency, and although ΔT is higher for higher Tc materials, that does not help if one wants a device that cools around room temperature or below.

The EC strength is defined as the ratio of the temperature change $\Delta T$ over the external electric field change $\Delta E$. Figure 7(b) shows the corresponding EC strength versus different compositions at the electric field difference $\Delta E_3$ = 150 MV/m and 200 MV/m. The maximum EC strengths for both $\Delta E_3$ occur at the MPB region with the composition $x$ = 0.45 and then the composition $x$ = 0.4. The appearance of the maximum EC effect at the composition $x$ = 0.45 within the MPB region is caused by the electric field–induced phase transition and easy rotation of the polarization under an external electric field near room temperature. The EC strength of the composition on the left of the MPB region is lower than that within the MPB region, which can be explained by the lower switchability of the polarization vector on the left of the MPB region than that within the MPB region. On the right of the MPB region is the tetragonal structure under all external electric fields. For the tetragonal structure with the $c$ axis along the [001] crystallographic direction and external electric field along the [001] direction, no polarization rotation occurs when the external electric field changes. Therefore, the EC strength on the right of the MPB region is lower than that within the MPB region. Finally, the minimum EC strength occurs at the composition $x$ = 1.0. The reported EC strengths in experiments are also plotted in Figure 7(b), which corresponds to the right side of the axis. The EC strength from our simulation is lower than that of the experimental observation.[58-60] Many factors contribute to this difference, such as the smaller simulation cell without a domain wall and the defect-free simulation model, among others.

## Conclusion

We find a set of phase transitions with increased electric field, which significantly affect the electrocaloric coefficients. We also find maximal ATC ΔT at 300 K for compositions within the MPB region $x$ = 0.45. A giant EC strength is found in the MPB region, where the polarization can easily rotate under a finite external electric field. For the concentration above $x = 0.50$, the solid solutions are in the tetragonal phase below $T_C$, where the behavior is similar to that of $PbTiO_3$.

## Acknowledgments

The authors gratefully acknowledge the 283  Gauss Centre for Supercomputing e.V. (www.gauss-centre.eu) for funding this project by 284  providing computing time on the GCS Supercomputer SuperMUC at Leibniz Supercom285 putting Centre (LRZ, www.lrz.de). This work was supported by the European Research Council under the Advanced Grant ToMCaT (Theory of Mantle, Core, and Technological Materials) and by the Carnegie Institution for Science. The authors appreciate their fruitful discussions with Dr. Shi Liu and Dr. Yangzheng Lin.

## References

[1]     J. Childress, Journal of Applied Physics **33**, 1793 (1962).
[2]     J. Scott, science **315**, 954 (2007).
[3]     S. Fähler *et al.*, Advanced Engineering Materials **14**, 10 (2012).
[4]     M. E. Lines and A. M. Glass, *Principles and applications of ferroelectrics and related materials* (Oxford university press, 1977).
[5]     G. G. Wiseman and J. K. Kuebler, Physical Review **131**, 2023 (1963).
[6]     A. Mischenko, Q. Zhang, J. Scott, R. Whatmore, and N. Mathur, Science **311**, 1270 (2006).
[7]     B. Neese, B. Chu, S.-G. Lu, Y. Wang, E. Furman, and Q. Zhang, Science **321**, 821 (2008).
[8]     H. Chen, T.-L. Ren, X.-M. Wu, Y. Yang, and L.-T. Liu, Applied Physics Letters **94**, 2902 (2009).
[9]     J. Hagberg, A. Uusimäki, and H. Jantunen, Applied Physics Letters **92** (2008).
[10]     X. Moya, E. Stern-Taulats, S. Crossley, D. González-Alonso, S. Kar-Narayan, A. Planes, L. Mañosa, and N. D. Mathur, Advanced Materials **25**, 1360 (2013).
[11]     F. Zhuo, Q. Li, J. Gao, Q. Yan, Y. Zhang, X. Xi, and X. Chu, Physical Chemistry Chemical Physics  (2017).
[12]     Y. Bai, X. Han, X.-C. Zheng, and L. Qiao, Scientific reports **3**, 2895 (2013).
[13]     Y. Bai, X. Han, and L. Qiao, Applied Physics Letters **102**, 252904 (2013).
[14]     M. C. Rose and R. E. Cohen, Physical review letters **109**, 187604 (2012).
[15]     M. Davis, D. Damjanovic, and N. Setter, Physical Review B **73**, 014115 (2006).
[16]     Z. Kutnjak, J. Petzelt, and R. Blinc, Nature **441**, 956 (2006).
[17]     Z. Kutnjak, R. Blinc, and Y. Ishibashi, Physical Review B **76**, 104102 (2007).
[18]     M. Liu and J. Wang, Scientific reports **5** (2015).
[19]     Z. Luo, D. Zhang, Y. Liu, D. Zhou, Y. Yao, C. Liu, B. Dkhil, X. Ren, and X. Lou, Applied Physics Letters **105**, 102904 (2014).
[20]     H.-H. Wu, J. Zhu, and T.-Y. Zhang, Nano Energy **16**, 419 (2015).
[21]     A. Chauhan, S. Patel, and R. Vaish, Energy Technology **3**, 177 (2015).
[22]     F. Le Goupil, A.-K. Axelsson, M. Valant, T. Lukasiewicz, J. Dec, A. Berenov, and N. M. Alford,

Applied Physics Letters **104**, 222911 (2014).

[23]     H.-H. Wu, J. Zhu, and T.-Y. Zhang, RSC Advances **5**, 37476 (2015).

[24]     F. Le Goupil, R. McKinnon, V. Koval, G. Viola, S. Dunn, A. Berenov, H. Yan, and N. M. Alford, Scientific Reports **6** (2016).

[25]     F. Weyland, M. Acosta, J. Koruza, P. Breckner, J. Rödel, and N. Novak, Advanced Functional Materials  (2016).

[26]     G. G. Guzmán-Verri and P. B. Littlewood, APL Materials **4**, 064106 (2016).

[27]     H.-H. Wu, J. Zhu, and T.-Y. Zhang, Physical Chemistry Chemical Physics **17**, 23897 (2015).

[28]     Z. Liu, X. Li, and Q. Zhang, Applied Physics Letters **101**, 082904 (2012).

[29]     S. Park and T. R. Shrout, Journal of Applied Physics **82** (1997).

[30]     B. Meyer and D. Vanderbilt, Physical Review B **65**, 104111 (2002).

[31]     H. Fu and R. E. Cohen, Nature **403**, 281 (2000).

[32]     M. Marathe, D. Renggli, M. Sanlialp, M. O. Karabasov, V. V. Shvartsman, D. C. Lupascu, A. Grünebohm, and C. Ederer, arXiv preprint arXiv:1703.05515  (2017).

[33]     M. Marathe, A. Grünebohm, T. Nishimatsu, P. Entel, and C. Ederer, Physical Review B **93**, 054110 (2016).

[34]     A. Grünebohm and T. Nishimatsu, Physical Review B **93**, 134101 (2016).

[35]     M. Sepliarsky, A. Asthagiri, S. Phillpot, M. Stachiotti, and R. Migoni, Current Opinion in Solid State and Materials Science **9**, 107 (2005).

[36]     Y. Zhang, J. Hong, B. Liu, and D. Fang, Nanotechnology **21**, 015701 (2009).

[37]     J. M. Vielma and G. Schneider, Journal of Applied Physics **114**, 174108 (2013).

[38]     O. Gindele, A. Kimmel, M. G. Cain, and D. Duffy, The Journal of Physical Chemistry C **119**, 17784 (2015).

[39]     Y.-B. Ma, A. Grünebohm, K.-C. Meyer, K. Albe, and B.-X. Xu, Physical Review B **94**, 094113 (2016).

[40]     N. Khakpash, H. Khassaf, G. Rossetti Jr, and S. Alpay, Applied Physics Letters **106**, 082905 (2015).

[41]     H. Khassaf, J. Mantese, N. Bassiri-Gharb, Z. Kutnjak, and S. Alpay, Journal of Materials Chemistry C **4**, 4763 (2016).

[42]     B. Dick Jr and A. Overhauser, Physical Review **112**, 90 (1958).

[43]     M. Sepliarsky, Z. Wu, A. Asthagiri, and R. Cohen, Ferroelectrics **301**, 55 (2004).

[44]     S. Tinte, B. Burton, E. Cockayne, and U. Waghmare, Physical review letters **97**, 137601 (2006).

[45]     B. P. Burton, E. Cockayne, and U. V. Waghmare, Physical Review B **72**, 064113 (2005).

[46]     P. Ganesh, E. Cockayne, M. Ahart, R. E. Cohen, B. Burton, R. J. Hemley, Y. Ren, W. Yang, and Z.-G. Ye, Physical Review B **81**, 144102 (2010).

[47]     P. Davies and M. Akbas, Journal of Physics and Chemistry of Solids **61**, 159 (2000).

[48]     M. Sepliarsky and R. E. Cohen, Journal of Physics: Condensed Matter **23**, 435902 (2011).

[49]     W. Smith and T. Forester, Journal of molecular graphics **14**, 136 (1996).

[50]     A. Asthagiri, Z. Wu, N. Choudhury, and R. E. Cohen, Ferroelectrics **333**, 69 (2006).

[51]     P. Mitchell and D. Fincham, Journal of Physics: Condensed Matter **5**, 1031 (1993).

[52]     Y. Liu, J. F. Scott, and B. Dkhil, Applied Physics Reviews **3**, 031102 (2016).

[53]     M. Ahart *et al.*, Nature **451**, 545 (2008).

[54]     W. Geng, Y. Liu, X. Meng, L. Bellaiche, J. F. Scott, B. Dkhil, and A. Jiang, Advanced Materials **27**, 3165 (2015).

[55]     F. Zhuo, Q. Li, J. Gao, Y. Wang, Q. Yan, Y. Zhang, X. Xi, X. Chu, and W. Cao, Applied Physics Letters **108**, 082904 (2016).

[56]     J. Peräntie, H. Tailor, J. Hagberg, H. Jantunen, and Z.-G. Ye, Journal of Applied Physics **114**, 174105 (2013).

[57]     M. C. Rose and R. E. Cohen, Physical Review Letters **112**, 249901 (2014).

[58]     L. Luo, H. Chen, Y. Zhu, W. Li, H. Luo, and Y. Zhang, Journal of Alloys and Compounds **509**, 8149 (2011).

[59]     L. Luo, M. Dietze, C.-H. Solterbeck, M. Es-Souni, and H. Luo, Applied Physics Letters **101**, 062907 (2012).

[60]    F. L. Goupil, A. Berenov, A.-K. Axelsson, M. Valant, and N. M. Alford, Journal of Applied Physics **111**, 124109 (2012).

Table 1. Shell model parameters of PMN-PT based on the first-principles calculation results within the local density approximation.

| Atom | Core charge | Shell charge | $k_2$ | $k_4$ |
|------|------------|--------------|-------|-------|
| Pb | 5.1471 | −3.3506 | 75.23 | 26915.85 |
| Ti | 9.6928 | −6.8081 | 1939.87 | 961.16 |
| Mg | 2.4533 | −0.1144 | 79.85 | 0.00 |
| Nb | 5.3086 | −2.1510 | 687.04 | 3983.87 |
| O | 0.7054 | −2.2658 | 25.83 | 1328.17 |
| Short range | $A$ | $B$ | $\rho$ | |
| Pb-O | 6286.380 | 296.2815 | 0.265236 | |
| Ti-O | 1409.971 | 4.8309 | 0.290702 | |
| Mg-O | 1039.504 | 63.3175 | 0.315801 | |
| Nb-O | 1507.531 | 4.0396 | 0.298814 | |
| O-O | 283.697 | −103.1517 | 0.520682 | |

Note: Core and shell charges are in units of electrons, energy in units of ev, and the length in unit of Å.

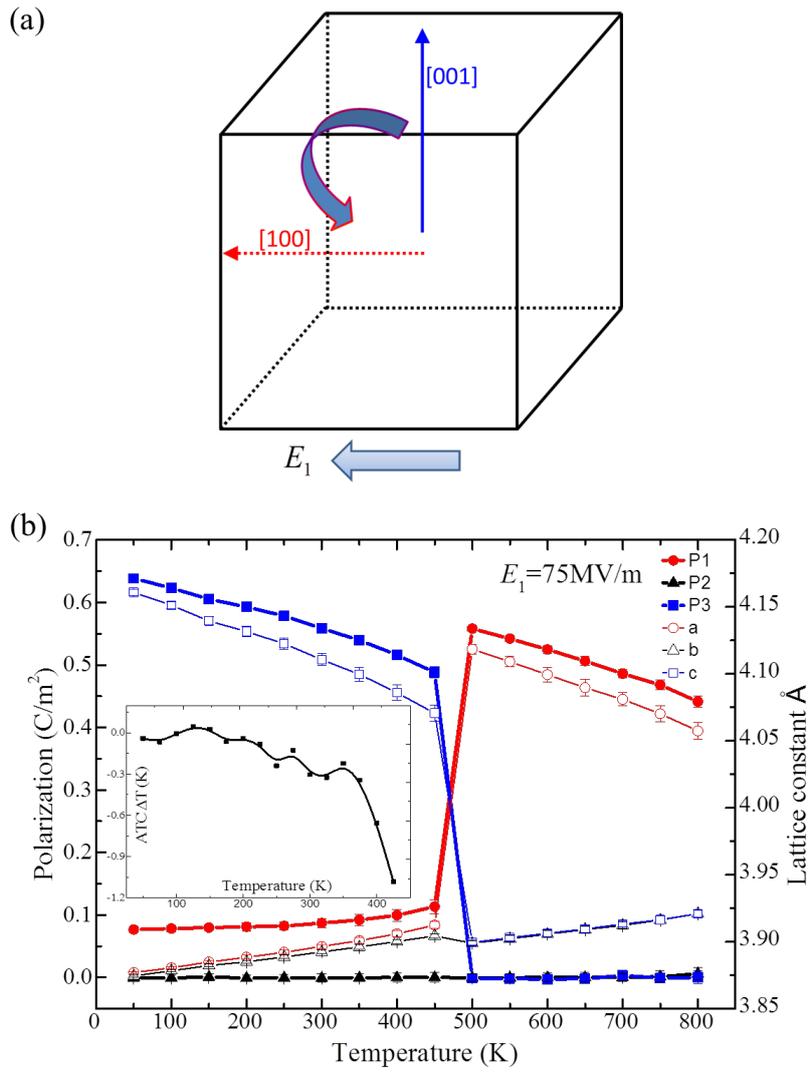

Figure 1. The [100] direction response of a [001]-polarized T phase, (a) the schematic of the [100] response for the [001]-polarized tetragonal phase, (b) the polarization component and lattice constant vs. the temperature of the composition PMN-0.8 PT.

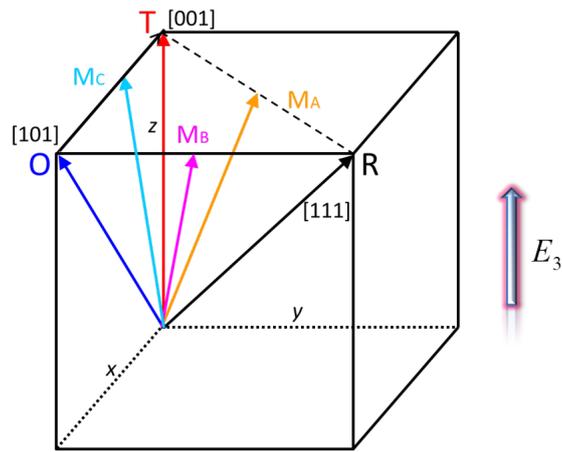

Figure 2. Schematic of the three types of monoclinic phases: $M_A$, $M_B$, and $M_C$ among T, R, and O phases in PMN-PT.

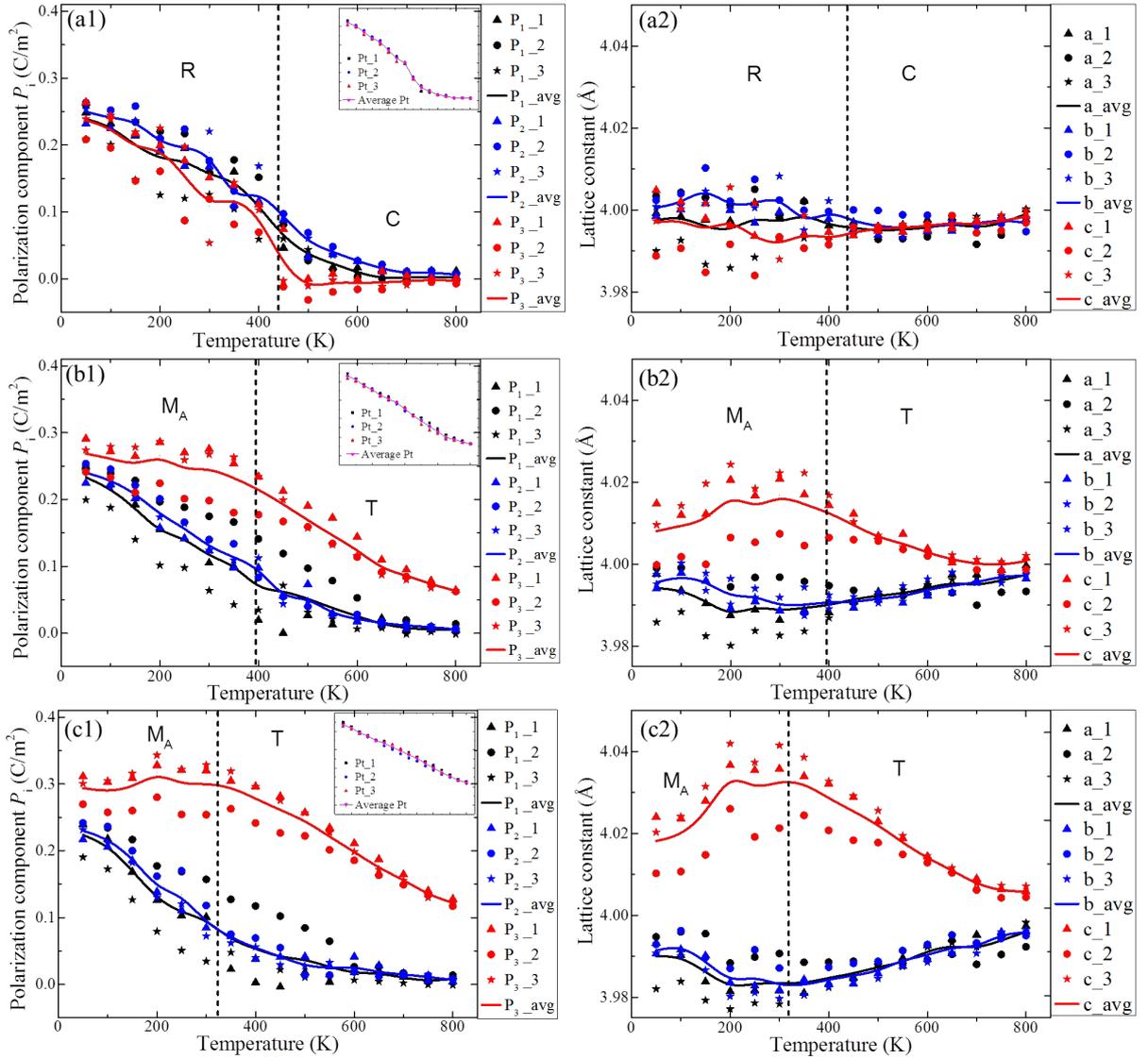

Figure 3. Average polarization component $P_i$ and lattice constant versus temperature of the PMN-0.3 PT on the left of the MPB region with respect to different external electric fields (a1)–(a2) $E_3 = 0$ MV/m, (b1)–(b2) $E_3 = 25$ MV/m, and (c1)–(c2) $E_3 = 50$ MV/m. The solid triangle, circle, and star symbols represent different layouts, and the solid lines link averages over configurations. The insets show the corresponding total polarization vs temperature with the same axis scale.

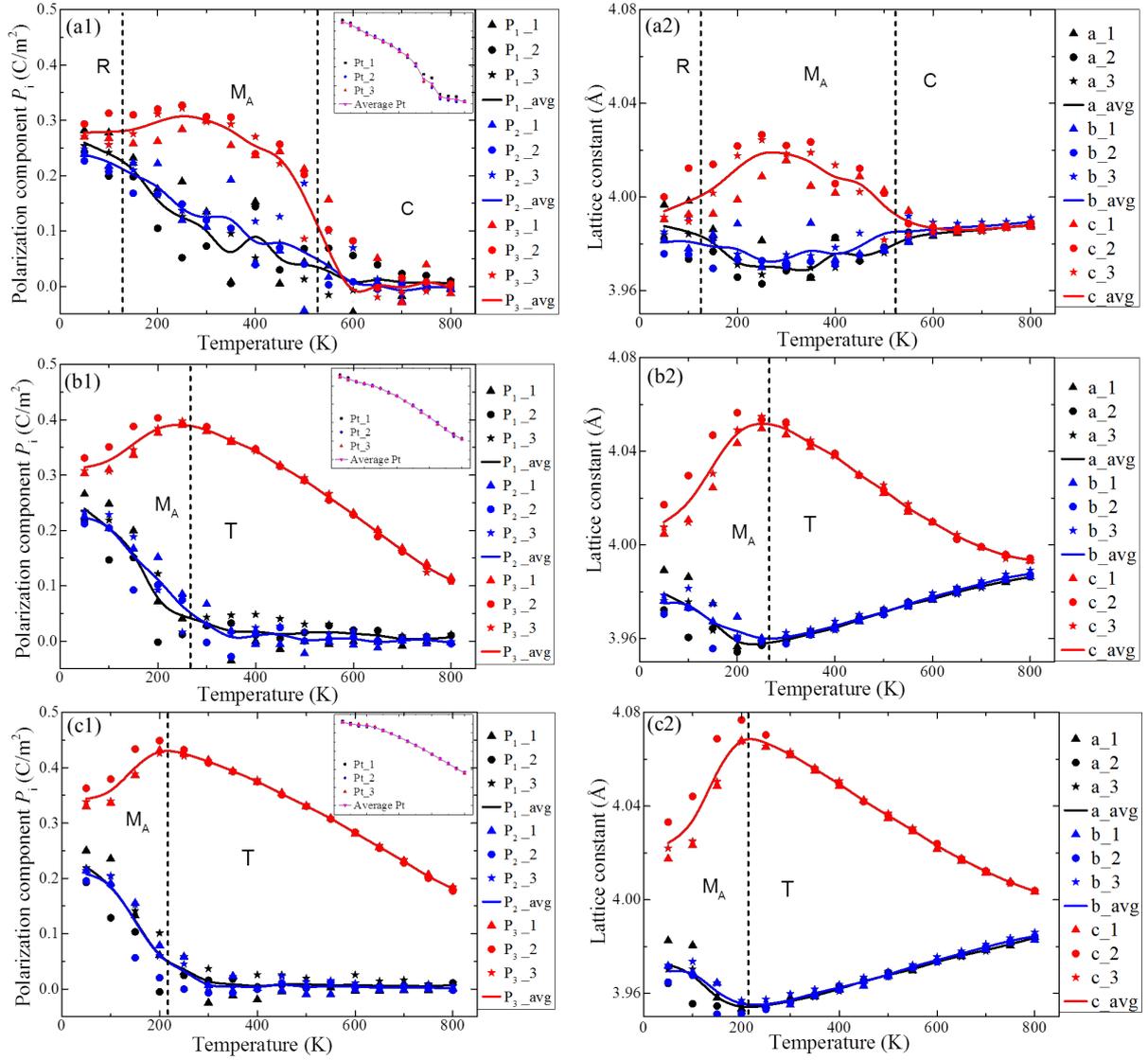

Figure 4. Average polarization component $P_i$ and lattice constant versus temperature of the PMN-0.4 PT with respect to different external electric fields (a1)–(a2) $E_3 = 0$ MV/m, (b1)–(b2) $E_3 = 25$ MV/m, and (c1)–(c2) $E_3 = 50$ MV/m. The solid triangle, circle, and star symbols represent different layouts, and the solid lines link averages over configurations. The insets show the corresponding total polarization vs temperature with the same axis scale.

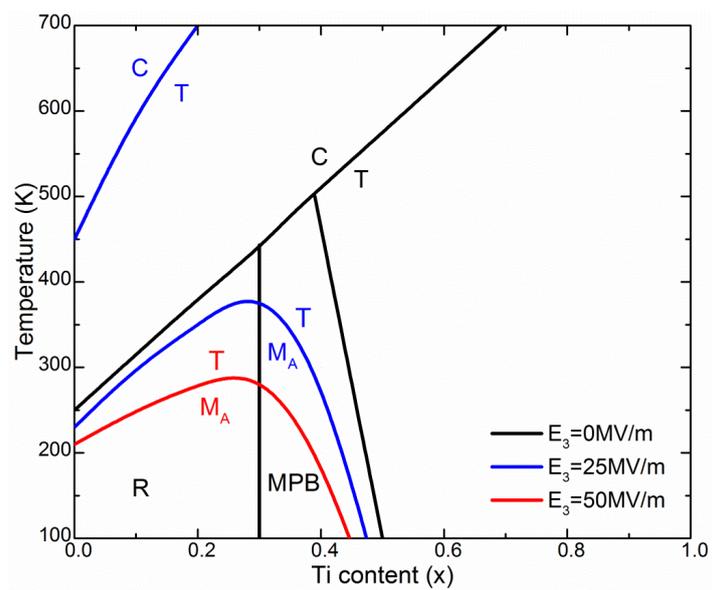

Figure 5. Temperature-composition phase diagram with respect to the different electric fields of PMN-PT.

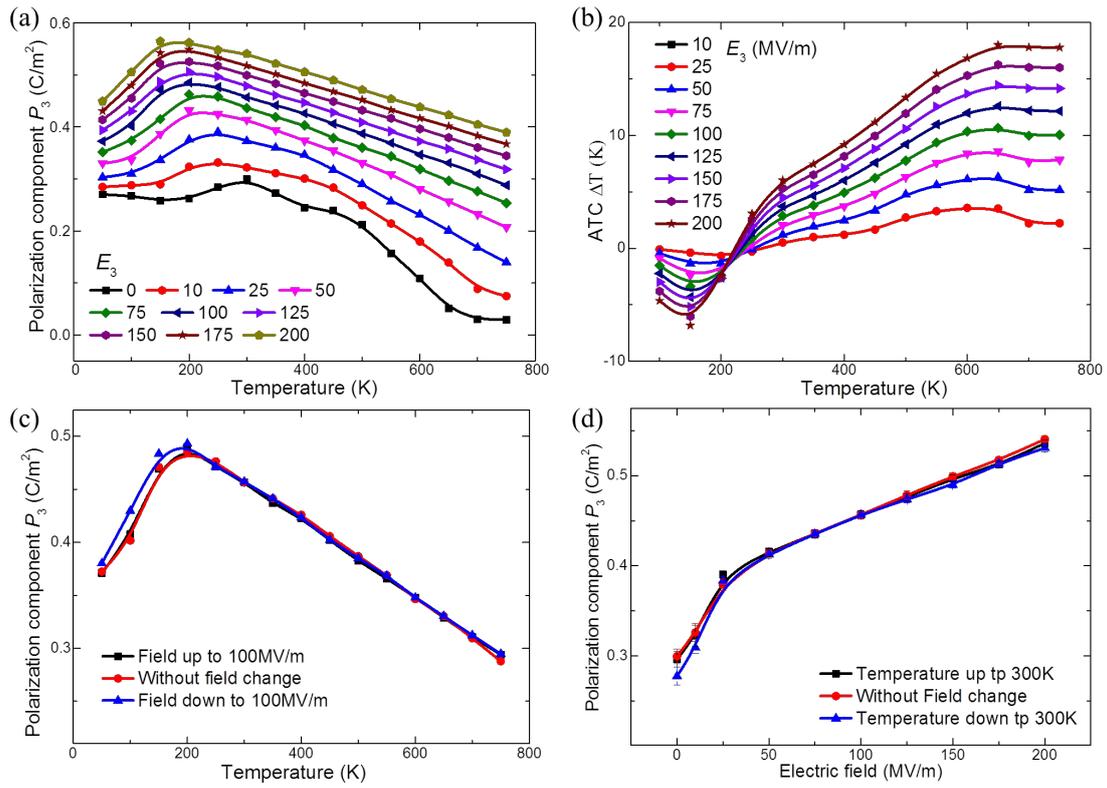

Figure 6. (a) Polarization component $P_3$ and (b) ATC $\Delta T$ versus temperature of the component $x = 0.4$; reversibility of (c) electric field 100 MV/m and (d) temperature 300 K–dependent polarization component $P_3$ of the composition $x = 0.4$.

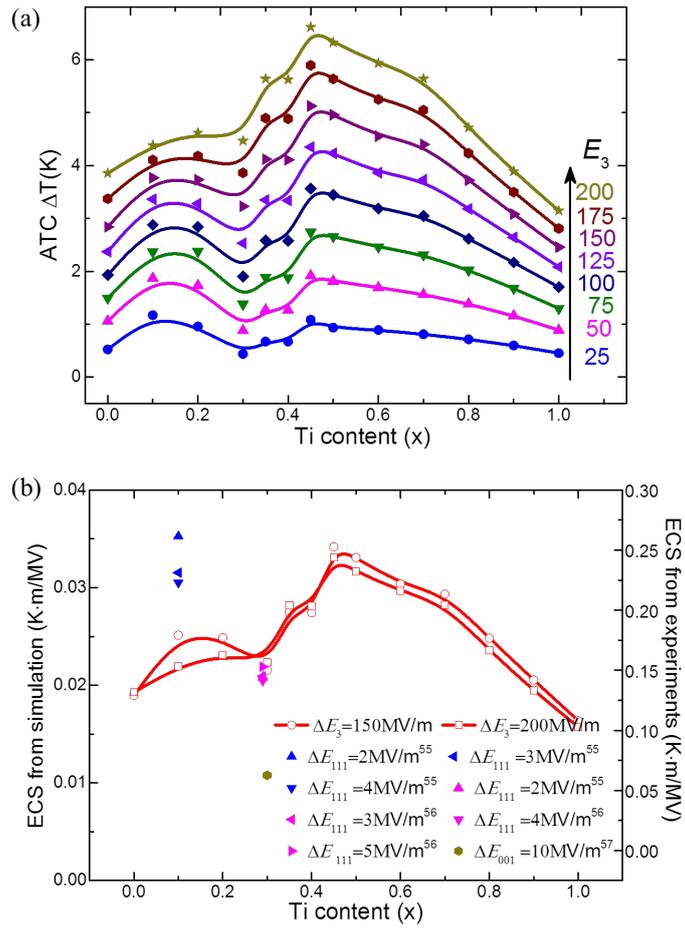

Figure 7. (a) ATC $\Delta T$ and (b) EC strength versus Ti content ($x$ = 0 to 1.0) from the averaged three layouts and experiments, where different color symbols correspond to different literature.